\documentclass[copyright,creativecommons]{eptcs}
\providecommand{\event}{CCA 2010}

\usepackage{amsmath, amsthm, amssymb}
\usepackage{latexsym}
\usepackage{amsfonts}

\title{The descriptive set-theoretic complexity of the set of points of continuity of a multi-valued function \\ (Extended Abstract)}

\author{Vassilios Gregoriades\thanks{The author is currently a post-doctoral researcher at the Mathematics Department of TU Darmstadt
in the work-group of \textsc{Ulrich Kohlenbach} (TU Darmstadt),
whom the author would like to thank. The author owns special
thanks to \textsc{Martin Ziegler} (TU Darmstadt) for bringing to
the author's attention the problem which motivated this article
and for his valuable advice. Finally the author would like to thank the referees of this article for their kind comments and suggestions.} \\
\institute{Technische Universit\"{a}t Darmstadt\\ GERMANY}
\email{gregoriades@mathematik.tu-darmstadt.de} }

\def\titlerunning{Descriptive Set-Theoretic complexity}
\def\authorrunning{V. Gregoriades}

\date{\today}

\begin{document}

\maketitle

\newcommand{\set}[2]{\ensuremath{\{#1 \ \mid \ #2\}}}

\newcommand{\om}{\ensuremath{\omega}}
\newcommand{\ep}{\ensuremath{\varepsilon}}

\newcommand{\iin}{\ensuremath{i \in \om }}
\newcommand{\n}{\ensuremath{n \in \om }}
\newcommand{\s}{\ensuremath{s \in \om }}

\newcommand{\ca}[1]{\ensuremath{ \mathcal{#1} }}

\newcommand{\del}{\ensuremath{\Delta^1_1 }}
\newcommand{\si}{\ensuremath{\Sigma^1_1 }}
\newcommand{\pe}{\ensuremath{\Pi^1_1 }}

\newcommand{\settwo}[2]{\ensuremath{ \set{#1_{#2}}{#2 \in \om} }}
\newcommand{\seq}[2]{\ensuremath{ (#1_{#2})_{#2 \in \omega} }}

\newcommand{\eq}{\ensuremath{\Longleftrightarrow }}

\newcommand{\arr}[1]{\ensuremath{ \overrightarrow{#1} }}
\newcommand{\barr}[1]{\ensuremath{ \overline{#1} }}

\newcommand{\gen}[1]{\ensuremath{ \ca{X}(#1) }}

\newcommand{\G}{\ensuremath{\Gamma }}

\newcommand{\empt}{\ensuremath{ \langle \cdot \rangle }}

\newcommand{\bottom}{\ensuremath{ -1 }}

\newcommand{\spat}[2]{\ensuremath{\ca{X}({#2})}}

\newcommand{\vg}[1]{}

\newcommand{\cod}{\ensuremath{\ca{T}}}

\newcommand{\omseq}{\ensuremath{\om^{\star}}}

\newtheorem{theorem}{Theorem}[section]
\newtheorem{lemma}[theorem]{Lemma}
\newtheorem{defn}[theorem]{Definition}
\newtheorem{prop}[theorem]{Proposition}
\newtheorem{cor}[theorem]{Corollary}
\newtheorem{rem}[theorem]{Remark}
\newtheorem{rems}[theorem]{Remarks}
\newtheorem{examp}[theorem]{Example}
\newtheorem{examps}[theorem]{Examples}

\begin{abstract}
In this article we treat a notion of continuity for a multi-valued
function $F$ and we compute the descriptive set-theoretic
complexity of the set of all $x$ for which $F$ is continuous at
$x$. We give conditions under which the latter set is either a
$G_\delta$ set or the countable union of $G_\delta$ sets. Also we
provide a counterexample which shows that the latter result is
optimum under the same conditions. Moreover we prove that those
conditions are necessary in order to obtain that the set of points
of continuity of $F$ is Borel i.e., we show that if we drop some
of the previous conditions then there is a multi-valued function
$F$ whose graph is a Borel set and the set of points of continuity
of $F$ is not a Borel set. Finally we give some analogue results
regarding a stronger notion of continuity for a multi-valued
function. This article is motivated by a question of M. Ziegler in
[{\em Real Computation with Least Discrete Advice: A Complexity
Theory of Nonuniform Computability with Applications to Linear
Algebra}, {\sl submitted}].
\end{abstract}

\section{Introduction.}

\label{section introduction}

A \emph{multi-valued} function $F$ from a set $X$ to another set
$Y$ is any function from $X$ to the power set of $Y$ i.e., $F$
assigns sets to points. Such a function will be denoted by $F:X
\Rightarrow Y$. A multi-valued function $F: X \Rightarrow Y$ can
be identified with its \emph{graph} $Gr(F) \subseteq X \times Y$
which is defined by
\[
(x,y) \in Gr(F) \eq y \in F(x).
\]
This way we view $F$ as a subset of $X \times Y$. From now on we
assume that all given multi-valued functions are between metric
spaces and that they are \emph{total} i.e., if $F : X \Rightarrow
Y$ is given then $F(x) \neq \emptyset$ for all $x \in X$, in other
words the projection of $F$ along $Y$ is the whole space $X$.

There are various notions of continuity for multi-valued
functions, here we focus on two of those (see
\cite{brattka-hertling} Definition 2.1,
\cite{choquet_multi_valued_functions} pp. 70-71 and
\cite{adamovicz_multivalued_functions} p. 82, p. 93).

\begin{defn}

\normalfont

\label{definition of continuity of multivalued functions}

Let $(X,p)$ and $(Y,d)$ be metric spaces; a multi-valued function
$F: X \Rightarrow Y$ is \emph{continuous at $x$} if there is some
$y \in F(x)$ such that for all $\ep > 0$ there is some $\delta >0$
such that for all $x' \in B_p(x,\delta)$ there is some $y' \in
F(x')$ for which we have that $d(y,y') < \ep$.
\end{defn}

\begin{defn}

\normalfont

\label{definition of strong continuity of multivalued functions}

Let $(X,p)$ and $(Y,d)$ be metric spaces; a multi-valued function
$F: X \Rightarrow Y$ is \emph{strongly continuous at $x$} if for
all $y \in F(x)$ and for all $\ep > 0$ there is some $\delta
>0$ such that for all $x' \in B_p(x,\delta)$ there is some $y' \in
F(x')$ for which we have that $d(y,y') < \ep$.
\end{defn}

It is clear that both these notions generalize the classical
notion of continuity of functions. Moreover it is also clear that
the continuity/strong continuity of a multi-valued function is
preserved under distance functions which generate the same
topology.

The motivation of this article is the following question posed by
M. Ziegler in \cite{martin} (Question 59(a)). It is well known that if we have a function $f: (X,p) \to (Y,d)$ then the set of points of continuity of $f$ is a $G_\delta$ subset of $X$; see for example 3.B in \cite{kechris}. So what can be said about the descriptive set-theoretic complexity of the set of points of continuity/strong continuity of a multi-valued function $F: (X,p) \Rightarrow (Y,d)$? In this article we present the answer for the case of continuity and some analogue results for the case of strong continuity. The full answer for the latter case is still under investigation.

We proceed with the basic terminology and notations. By \om \ we
denote the set of natural numbers (including the number $0$).
Suppose that $X$ and $Y$ are two topological spaces. We call a
function $f: X \to Y$ a \emph{topological isomorphism} between $X$
and $Y$ if the function $f$ is bijective, continuous and the
function $f^{-1}$ is continuous. We will also say that the space
$X$ is \emph{topologically isomorphic} with $Y$ is there exists a
topological isomorphism between $X$ and $Y$.

The \emph{Baire space} \ca{N} is the set of all sequences of
naturals i.e., $\ca{N} = \om^\om$ with the usual product topology.
We call the members of the Baire space as \emph{fractions} and we
usually denote them by lower case Greek letters $\alpha, \beta$
etc. One choice of basic neighborhoods for the product topology on
\ca{N} is the collection of the following sets
$$
N(k_0,\dots,k_{n-1}) = \set{\alpha \in \ca{N}}{\alpha(0)=k_0,
\dots, \alpha(n-1) = k_{n-1}}
$$
where $k_0,\dots,k_{n-1} \in \om$. The set of ultimately constant
sequences  is clearly countable and dense in \ca{N}; thus the
latter is a separable space. For $\alpha, \beta \in \ca{N}$ with
$\alpha \neq \beta$ define
$$
d_{\ca{N}}(\alpha,\beta) = 1 / (\textrm{least} \ n \ [\alpha(n)
\neq \beta(n)] \ + 1).
$$
Also put $d_{\ca{N}}(\alpha,\alpha) =0$ for all $\alpha \in
\ca{N}$. It is not hard to see that the function $d_{\ca{N}}$ is a
complete distance function on \ca{N} which generates its topology.
From now on we think of the Baire space \ca{N} with this distance
function $d_{\ca{N}}$.

We denote by \ca{C} the subset of the Baire space \ca{N} which
consist of all sequences which values $0$ and $1$ i.e., $\ca{C} =
2^\om$. The set \ca{C} with the induced topology is a compact
space. It is not hard to see that \ca{C} is topologically
isomorphic with the usual Cantor set of the unit interval. This
result motivates us to call \ca{C} as the \emph{Cantor space}.

We denote by \omseq \ the set of all finite sequences of \om. If
$u \in \omseq$ then there are unique naturals
$n,k_0,\dots,k_{n-1}$ such that $u = (k_0,\dots,k_{n-1})$. The
\emph{length of $u$} is the previous natural $n$ and we denote it
by $lh(u)$. Also we write $u(i) = k_i$ for all $i < lh(u)$, so
that $u=(u(0),\dots,u(lh(u)-1))$. It is convenient to include the
\emph{empty sequence} in \omseq \ i.e., the one with zero length.
The latter will be denoted by \empt. So when we write
$u=(u(0),\dots,u(n-1))$ we will always mean in case where $n=0$
that $u = \empt$. If $u \in \omseq$ and $n \in \om$ we denote the
finite sequence $(u(0),\dots,u(lh(u)-1),n)$ by $u \ \hat{} \ (n)$.
We write $u \sqsubseteq v$ exactly when $lh(u) \leq lh(v)$ and
$u(i) = v(i)$ for all $i < lh(u)$ i.e., $u \sqsubseteq v$ means
that $v$ is an extension of $u$ or equivalently $u$ is an initial
segment of $v$.

A set $T \subseteq \omseq$ is called a \emph{tree} on \om \ if it
is closed under initial segments i.e.,
$$
v \in T \ \& \ u \sqsubseteq v \ \ \Longrightarrow \ \ u \in T.
$$
The members of a tree $T$ are called \emph{nodes} or
\emph{branches} of $T$. A tree $T$ is of \emph{finite branching}
if and only if for all $u \in T$ there are only finitely many $n
\in \om$ such that $u \ \hat{} \ (n) \in T$. A fraction $\alpha$
is an \emph{infinite branch of $T$} if and only if for all \n \ we
have that $(\alpha(0),\dots,\alpha(n-1)) \in T$. The \emph{body}
$[T]$ of a tree $T$ is the set of infinite branches of $T$.

For practical reasons when we refer to a tree $T$ we will always
assume that $T$ is not empty i.e., $\empt \in T$. Define
\[
Tr = \set{T \subseteq \omseq}{\textrm{the set $T$ is a tree on
\om}}.
\]
We may view every $T \in Tr$ as a member of $2^{\omseq}$ by
identifying $T$ with its characteristic function \newline $\chi_T : \omseq
\to \{0,1\}$. Since the set \omseq \ is countable the space
$2^{\omseq}$ with the product topology is completely metrizable -
in fact it is topologically isomorphic with the Cantor space
\ca{C}. Moreover the set $Tr$ is a closed subset of \ca{C}. Indeed
let $T_i \in Tr$ for all \iin \ be such that $T_i \stackrel{i \to
\infty}{\longrightarrow}S$ for some $S \in 2^{\omseq}$; we will
prove that $S \in Tr$. From the hypothesis it follows that for all
$u \in \omseq$ there is some $i_0 \in \om$ such that for all $i
\geq i_0$ we have that
\[
u \in T_i \eq u \in S.
\]
Taking $u = \empt$ since $T_i \in Tr$ for all \iin \ we have that
$\empt \in S$ and so $S$ is not empty. Also if $u, v \in \omseq$
we find $i$ large enough so that $u \in T_i \eq u \in S$ and $v
\in T_i \eq v \in S$. So if $u \in S$ and $v \sqsubseteq u$ then
$u \in T_i$ and since $T_i$ is a tree we also have that $v \in
T_i$; hence $v \in S$. Therefore $S \in Tr$ and the set of trees
$Tr$ is closed in \ca{C}.

We make a final comment about trees. For any non-empty set $S$ of
finite sequences of naturals the tree $T$ which is \emph{generated
by $S$} is the following
\[
\set{u}{(\exists w \in S)[ u \sqsubseteq w]}
\]
i.e., the tree which is generated by $S$ is the tree which arises
by taking all initial segments of members of $S$.

Suppose that $X$ is a metric space. The family $\Sigma^0_1(X)$ is
the collection of all \emph{open} subsets of $X$. Inductively we
define the family $\Sigma^0_{n+1}(X)$ for $n \geq 1$as follows:
for $A \subseteq X$,
\[
A \in \Sigma^0_{n+1}(X) \ \eq \ A = \bigcup\limits_{\iin} A_i, \ \
\ \textrm{where $X \setminus A_i \in \Sigma^0_{k_i}(X)$ for some
$k_i \leq n$ for all \iin.}
\]
Put also
\[
\Pi^0_n(X) =\set{B \subseteq X}{X \setminus B \in \Sigma^0_n(X)}
\]
and $\Delta^0_n(X) = \Sigma^0_n(X) \cap \Pi^0_n(X)$ for all $n
\geq 1$. Notice that family $\Pi^0_1(X)$ is the collection of all
closed subsets of $X$, the family $\Sigma^0_2(X)$ is the
collection of all $F_\sigma$ subset of $X$ and so on. By a simple
induction one can prove that $\Sigma^0_n(X) \cup \Pi^0_n(X)
\subseteq \Delta^0_{n+1}(X)$ for all $n \geq 1$. It is well known
that in case where $X$ admits a complete distance function and it
is an uncountable set then $\Sigma^0_n(X) \neq \Pi^0_n(X)$ for all
$n \geq 1$ and so the previous inclusion is a proper one for all
$n \geq 1$, (see \cite{kechris} and \cite{yiannis}).

The families $\Sigma^0_n(X), \Pi^0_n(X)$ are closed under finite
unions, finite intersections, and continuous pre-images i.e., if
$f: X \to Y$ is continuous and $B \subseteq Y$ is in
$\Sigma^0_n(Y)$ then $f^{-1}[B]$ is in $\Sigma^0_n(X)$. Moreover
it is clear that if $f: X \to Y$ is a topological isomorphism then
for all $A \subseteq X$ we have that $A \in \Sigma^0_n(X)$ if and
only if $f[A] \in \Sigma^0_n(X)$ and similarly for $\Pi^0_n(X)$
for all \n. Finally the family is $\Sigma^0_n(X)$ is closed under
countable unions, the family $\Pi^0_n(X)$ is closed under
countable intersections and the family $\Delta^0_n(X)$ is closed
under complements. We usually say that $A$ is in $\Sigma^0_n$ when
$X$ is easily understood from the context. It is clear that all
sets in $\Sigma^0_n$ are Borel sets.

We now deal with a bigger family of sets. Suppose that $X$ is
separable and that $X$ admits a complete distance function. A set
$A \subseteq X$ is in $\Sigma^1_1(X)$ or it is \emph{analytic} if
$A$ is the continuous image of a closed subset of a complete and
separable metric space.\footnote{The notion of an analytic set can
be treated in a more general context of spaces; however we prefer
to stay in the context of complete and separable metric spaces.}
It is well known that in the definition of analytic sets we may
replace the term ``continuous image" by ``Borel image" (i.e.,
image under a Borel measurable function) and/or the term ``closed
subset" by ``Borel set"", (see \cite{kechris} and \cite{yiannis}).
A set $B \subseteq X$ is in $\Pi^1_1(X)$ or it is
\emph{co-analytic} if the set $X \setminus B$ is analytic and $B$
is in $\Delta^1_1(X)$ or it is \emph{bi-analytic} if $B$ is both
analytic and co-analytic. It is well known that every Borel set is
analytic, hence every Borel set is bi-analytic. A classical
theorem of Suslin states that a set $B$ is Borel if and only if it is
bi-analytic, (see \cite{yiannis} 2E.1 and 2E.2).

The families $\Sigma^1_1(X), \ \Pi^1_1(X)$ and $\Delta^1_1(X)$ are
closed under countable unions, countable intersections and
continuous pre-images. Moreover the family $\Sigma^1_1(X)$ is
closed under Borel images i.e., if $Y$ is a complete and separable
metric space, $f: X \to Y$ is a Borel measurable function and $A$
is an analytic subset of $X$ then $f[A]$ is an analytic subset of
$Y$. Finally if $X$ is uncountable we have that $\Sigma^1_1(X)
\neq \Pi^1_1(X)$ and in particular there is an analytic set which
is not Borel.

We can pursue this hierarchy further by defining the family $\Sigma^1_{n+1}(X)$ as the collection of all subsets of $X$ which are the continuous image of a $\Pi^1_n$ subset of a complete and separable metric space. Similarly one defines the family $\Pi^1_{n+1}(X)$ as the collection of all subsets of $X$ whose complement is in $\Sigma^1_{n+1}(X)$ and the family $\Delta^1_{n+1}(X)$ as the collection of all subsets of $X$ which belong both to $\Sigma^1_{n+1}(X)$ and $\Pi^1_{n+1}(X)$. By a simple
induction one can prove that $\Sigma^1_n(X) \cup \Pi^1_n(X)
\subseteq \Delta^1_{n+1}(X)$ for all $n \geq 1$. Also the analogous properties stated above are true. The reader may refer to \cite{yiannis} for more information on those classes.

The proofs of the forthcoming theorems make a substantial use of
techniques of Descriptive Set Theory which involve the use of many
quantifiers. Of course those quantifiers can be interpreted as
unions and intersections of sets and this is what we usually do in
order to prove that a given set is for example $\Pi^0_2$. There
are some cases though (for example in the proof of Theorem
\ref{theorem counterexample about S-3}) where this interpretation
becomes too complicated. In these cases it is better to think of a
given set $P$ as a relation in order to derive its complexity. The
reader can consult section 1C of \cite{yiannis} on how one can
make computations with relations.

\section{Results about the set of points of continuity of a multi-valued function.}

\label{Results about the set of points of continuity}

We begin with some positive results regarding the set of points of
continuity of a multi-valued function $F$. Recall that a
topological space $Y$ is \emph{exhaustible by compact sets} if
there is a sequence $(K_n)_{\n}$ of compact subsets of $Y$ such
that every $K_n$ is contained in the interior of $K_{n+1}$ and $Y
= \bigcup\limits_{\n} K_n$. Notice the lack of any hypothesis
about the set $F$ in the next theorem.

\begin{theorem}

\label{theorem G-d when compact and Sigma-3 when sigma-compact}

Let $(X,p)$ and $(Y,d)$ be metric spaces with $(Y,d)$ being
separable and let $F: X \Rightarrow Y$ be a multi-valued function.

\begin{enumerate}

\item[(a)] If  the set $F(x)$ is compact for all $x \in X$ then the set of
points of continuity of $F$ is $\Pi^0_2$ i.e., $G_\delta$.

\item[(b)] If $Y$ is exhaustible by compact sets and
the set $F(x)$ is closed for all $x \in X$, then the set of points
of continuity of $F$ is $\Sigma^0_3$.
\end{enumerate}

\end{theorem}

Theorem \ref{theorem G-d when compact and Sigma-3 when
sigma-compact} has an interesting corollary which answers Question
59(a) posed by M. Ziegler in \cite{martin}.

\newpage

\begin{cor}

\label{corollary from theorem G-d and Sigma-3}

Suppose that $X$ is a metric space and that $F: X \Rightarrow
\mathbb{R}^m$ is a multi-valued function such that the set $F(x)$
is closed for all $x \in X$.

\item[(a)] The set of the points of continuity of $F$ is $\Sigma^0_3$.

\item[(b)] If moreover the set $F(x)$ is bounded for all $x \in X$ then the
set of points of continuity of $F$ is $\Pi^0_2$.
\end{cor}

We now show that the results of Theorem \ref{theorem G-d when
compact and Sigma-3 when sigma-compact} are optimum. It is well
known that there are functions $f: [0,1] \to \mathbb{R}$ for which
the set of points of continuity is not $F_\sigma$. Therefore the
$\Pi^0_2$-answer is the best one can get. Thus we only need to
deal with the $\Sigma^0_3$-answer. The following lemmas, although
being straightforward from the definitions, will prove an elegant
tool for the constructions that will follow.

\begin{lemma}

\label{lemma about extension}

Suppose that $(X_0,p_0)$, $(X_1,p_1)$, $(Y,d)$ are metric spaces
and that there exists a function $f: X_0 \to X_1$ such that $f[X_0]$ is closed and $f: X_0 \to f[X_0]$ is a topological isomorphism. Assume
that we are given a multi-valued function $F: X_0 \Rightarrow Y$.
Define the multi-valued function $\tilde{F}: X_1 \Rightarrow Y$ as
follows:
\[
\tilde{F}(x_1) = F(x_0) \ \ \ \textrm{if $x_1 = f(x_0)$ for some
$x_0 \in X_0$ and} \ \ \ \tilde{F}(x_1) = Y \ \ \
\textrm{otherwise}.
\]
Then
\begin{enumerate}

\item $\tilde{F}$ is continuous at $x_1$ if and only if either $x_1 \not
\in f[X_0]$ or $x_1=f(x_0)$ and $F$ is continuous at $x_0$. Hence
if we denote by $P_0$ and $P_1$ the set of points of continuity of
$F$ and $\tilde{F}$ respectively we have that
\[
P_1 = f[P_0] \cup (X_1 \setminus f[X_0]).
\]
\item If $\Gamma$ is any of the classes $\Sigma^0_n, \Pi^0_n$, with
$n\geq2$ or $\Delta^1_1$ then
\[
P_1 \in \Gamma \eq P_0 \in \Gamma.
\]
In particular if the set of points of continuity of $F$ is not
$\Pi^0_3$ (Borel) then the set of points of continuity of
$\tilde{F}$ is not $\Pi^0_3$ (Borel respectively).

Moreover the sets $F$ and $\tilde{F}$ as subsets of $X_0 \times Y$ and $X_1 \times Y$ respectively satisfy
\[
F \in \Gamma \eq \tilde{F} \in \Gamma.
\]

\item If $F(x_0)$ is a closed subset of $Y$ for all $x_0 \in X_0$
then $\tilde{F}(x_1)$ is also a closed subset of $Y$ for all $x_1
\in X_1$.
\end{enumerate}

\end{lemma}

Lemma \ref{lemma about extension} has a cute corollary which might
be regarded as the multi-valued analogue of the Tietze Extension
Theorem.

\begin{cor}

\label{corollary analogue of Tietze extensionality}

Every continuous multi-valued function which is defined on a
closed subset of a metric space can be extended continuously on
the whole space.

\end{cor}

\enlargethispage{5pt}

\begin{lemma}

\label{lemma about composition}

Let $X, Y, Z$ be metric spaces, $F: X  \Rightarrow Y$ be a multi-valued function and $\pi: Y \to Z$ be a topological isomorphism between $Y$ and $\pi[Y]$. Define the composition $\pi \circ F : X \Rightarrow Z:$
\[
(\pi \circ F)(x) = \pi[F(x)], \ \ x \in X.
\]
The following hold.
\begin{enumerate}

\item A point $x \in X$ is a point of continuity of $F$ if and only if $x$ is a point of continuity of $\pi \circ F$;

\item If the set $\pi[Y]$ is closed and the set $F(x)$ is closed for some $x \in X$ then the set $(\pi \circ F)(x)$ is also closed.

\item If $F$ is a Borel subset of $X \times Y$ then $\pi \circ F$ is a Borel subset of $X \times Z$.
\end{enumerate}

\end{lemma}

\begin{theorem}

\label{theorem counterexample about S-3}

There is a multi-valued function $F: [0,1] \to \mathbb{R}$ such
that the set $F(x)$ is closed for all $x$, the set $F$ is a
$\Pi^0_2$ subset of $[0,1] \times \mathbb{R}$ and the set of
points of continuity of $F$ is not $\Pi^0_3$. Therefore the
$\Sigma^0_3$-answer is the best possible for a multi-valued
function $F$ from $[0,1]$ to $\mathbb{R}$ even if $F$ is below the
$\Sigma^0_3$-level.
\end{theorem}

One can ask what is the best that we can say about the set of
points of continuity of $F$ without any additional topological
assumptions for $Y$ or for $F(x)$. The following proposition gives
an upper bound for the complexity of this set. (Notice though that
we restrict ourselves to complete and separable metric spaces.)

\begin{prop}

\label{proposition the set of points of continuity is analytic}

Let $(X,p)$ and $(Y,d)$ be complete and separable metric spaces
and let $F: X \Rightarrow Y$ be a multi-valued function such that the set $F \subseteq X \times Y$ is analytic. Then the set of points of continuity of $F$ is analytic as well.
\end{prop}

Now we show that if we remove just one of our assumptions about $F(x)$ or about $Y$ in Theorem \ref{theorem G-d when compact and Sigma-3 when
sigma-compact}, then it is possible that the set of
points of continuity of $F$ is not even a Borel set. Therefore
Proposition \ref{proposition the set of points of continuity is
analytic} is the best that one can say in the general case.

\begin{theorem}{ \ }

\label{theorem counterexample for analytic}

\begin{enumerate}

\item[(a)] There is a multi-valued function $F: \ca{C} \Rightarrow \ca{N}$
such that the set $F(x)$ is closed for all $x \in \ca{C}$ and the
set of points of continuity of $F$ is analytic and not Borel. Moreover the set $F$ is a Borel subset of $\ca{C} \times \ca{N}$.

\item[(b)] There is a multi-valued function $F: [0,1] \Rightarrow [0,1]$ for
which the set of the points of continuity of $F$ is analytic and not
Borel. Moreover the set $F$ is a Borel subset of $[0,1] \times [0,1]$.
\end{enumerate}

\end{theorem}

It is perhaps useful to make the following remarks. If we replace in (a) of Theorem \ref{theorem G-d when compact and Sigma-3 when sigma-compact} the condition about $F(x)$ being compact for all $x$  with ``$F(x)$ is closed for all
$x$", then from (a) of Theorem \ref{theorem counterexample for analytic} we can see that the result fails in the worst possible way. Also -in
connection with (b) of Theorem \ref{theorem G-d when compact and
Sigma-3 when sigma-compact}- we can see that if we drop the
hypothesis about $Y$ being exhaustible by compact sets but keep
the second condition ``$F(x)$ is closed for all $x$", then again
the result fails in the worst possible way.

If we replace in (a) of Theorem
\ref{theorem G-d when compact and Sigma-3 when sigma-compact} the hypothesis ``$F(x)$ is compact for all $x$", with
``$Y$ is compact" then still the result fails in the worst
possible way.

In conclusion if we want to obtain that
the set of points of continuity of a multi-valued function $F$ is
Borel, then we cannot drop the condition ``$F(x)$ is closed for
all $x$". But yet this condition alone is not sufficient in order to
derive this result as long as $Y$ is neither compact nor
exhaustible by compact sets.

Below we give a brief sketch of the proof of the latter theorem.

\emph{Sketch of the proof.}
Let $Tr$ be the set of all (non-empty) trees on
\om, (see the Introduction). As we mentioned before the set $Tr$
can be regarded as a compact subspace of the Cantor space \ca{C}.
From Lemma \ref{lemma about extension} it is enough to construct a
multi-valued function $F: Tr \Rightarrow \ca{N}$ such that the set
of points of continuity of $F$ is not Borel and the set $F(T)$ is
closed for all $T \in Tr$.

Denote by $IF$ the set of all ill founded trees i.e, the set of
all $T \in Tr$ for which the body $[T]$ is not empty. It is well
known (see \cite{kechris} 27.1) that the set $IF$ is an analytic subset of
$Tr$ which is not Borel.\footnote{A classical way for proving that
a given set $A \subseteq \ca{X}$ is not Borel is finding a Borel
function $\pi: Tr \to \ca{X}$ such that $IF = \pi^{-1}[A]$. If
$A$ was a Borel set then $IF$ would be Borel, a contradiction.} For
$T \in Tr$ we define the tree
\[
T^{+1} = \set{(u(0)+1, \dots, u(n-1)+1)}{u \in T, lh(u) = n}.
\]
Also we define the set $trm(T)$ as the set of all \emph{terminal}
nodes of $T$ i.e., the set of all those $u$'s in $T$ for which
there is no $w \in T$ such that $u \sqsubseteq w$ and $u \neq w$.
Define the multi-valued function $F: Tr \Rightarrow \ca{N}$ as
follows
\[
F(T) = [T^{+1}] \cup \set{u \ \hat{} \ (0,0,0,\dots)}{u \in
trm(T^{+1})}
\]
for all $T \in Tr$.

Then we prove that (1) the set $F$ is a Borel subset of $\ca{C} \times \ca{N}$, (2) for all $T \in Tr$ the set $F(T)$ is a
closed subset of the Baire space \ca{N} and (3) the multi-valued function $F$ is continuous at $T$ if and only if $T \in IF$. The second assertion of the theorem follows from the first one and Lemmas \ref{lemma about extension} and \ref{lemma about composition}.

\textbf{Question 1.} \normalfont Suppose that we are given a
multi-valued function $F: X \Rightarrow Y$ for which we have that
the set $F(x)$ is closed for all $x$ and $Y$ is separable. As we
have proved before in case where $Y$ is exhaustible by compact
sets the set of points of continuity of $F$ is $\Sigma^0_3$ and in
case where $Y = \ca{N}$ it is possible that the latter set is not
even Borel. In fact one can see that the latter is true not just
for $Y = \ca{N}$ but also in case where \ca{N} is topologically
isomorphic with a closed subset of $Y$.
The question is what happens when $Y$ falls in neither of the
previous cases i.e., $Y$ is neither exhaustible by compact sets
nor it contains \ca{N} as a closed subset. An interesting class of
such examples is the class of infinite dimensional separable
\emph{Banach spaces} i.e., (infinite dimensional) linear normed
spaces which are complete and separable under that norm. Any such
space is not exhaustible by compact sets and it does not contain
\ca{N} as a closed subset. Therefore the theorems of this article
provide no information in this case. It would be interesting to
find the best upper bound for the complexity of the set of points
of continuity of $F$ when $Y$ is an infinite dimensional separable
Banach space and the set $F(x)$ is closed for all $x$.

\section{Strong Continuity.}

\label{section strong continuity}

We continue with some results regarding the set of
points of \emph{strong} continuity of a multi-valued function $F$.
In particular we will prove the corresponding of Theorem
\ref{theorem G-d when compact and Sigma-3 when sigma-compact} and
Proposition \ref{proposition the set of points of continuity is
analytic}. The existence of examples which show that these results are optimum is still a subject under investigation.

Let us begin with some remarks. As we mentioned in the beginning,
Theorem \ref{theorem G-d when compact and Sigma-3 when
sigma-compact} does not require any additional hypothesis about
$F$ as a subset of $X \times Y$. However the following remark
suggests that this is not the case for strong continuity.

\begin{rem}

\label{remark about strong continuity}

\normalfont

Let $A$ be a dense subset of $[0,1]$; define the multi-valued function $F: [0,1] \Rightarrow \{0,1\}$ as follows
\[
F(x) = \{0\}, \ \ \textrm{if $x \in A$ and} \ F(x) = \{0,1\} \ \ \textrm{if $x \not \in A$},
\]
for all $x \in [0,1]$. We claim that the set of points of strong continuity of $F$ is exactly the set $A$. Let $x \in A$, $y \in F(x)$ and $\ep > 0$. Take $\delta = 1 > 0$ and let $x' \in (x-\delta,x+\delta)$. We have that $y = 0$ and also since $0 \in F(x')$ we can take $y' = 0$; so $|y-y'| = 0 < \ep$. Now let $x \not \in A$. We take $y = 1 \in F(x)$ and $\ep = \frac{1}{2}$. Let any $\delta >0$. Since $A$ is a dense subset of $[0,1]$ there is some $x' \in A$ such that $x' \in (x-\delta,x+\delta)$. Clearly for all $y' \in F(x')$ we have that $y' = 0$ and so $|y - y'| = 1 > \ep$.
\end{rem}

Since there are dense subsets of $[0,1]$ which are way above the level of analytic sets from Remark \ref{remark about strong continuity} we can see that there is no hope to obtain the corresponding of Theorem \ref{theorem G-d when compact and Sigma-3 when sigma-compact} without any additional assumptions about the complexity of the set $F$. Notice also that those assumptions about the set $F$ have to be at least as strong as the result that we want to derive. For example it is well known that there is a dense $\Pi^0_3$ set $A \subseteq [0,1]$ which is not $\Sigma^0_3$;  hence by taking the multi-valued function $F$ of Remark \ref{remark about strong continuity} with respect to that set $A$ we can see that $F$ is $\Delta^0_4$ as a subset of $[0,1] \times [0,1]$ and that the set of points of strong continuity of $F$ (i.e., the set $A$) is not $\Sigma^0_3$. In other words if we want to result to a $
\Sigma^0_3$ set we need to assume that $F$ does not go above the third level of the Borel hierarchy. The following may be regarded as the corresponding strong-continuity analogue of Theorem \ref{theorem G-d when compact and Sigma-3 when sigma-compact}.

\begin{theorem}

\label{theorem strong continuity G-d when compact and Sigma-3 when sigma-compact}

Let $(X,p)$ and $(Y,d)$ be metric spaces with $(Y,d)$ being
separable and let $F: X \Rightarrow Y$ be a multi-valued function such that $F$ is a $\Sigma^0_2$ subset of $X \times Y$.

\begin{enumerate}

\item[(a)] If $Y$ is compact and the set $F(x)$ is closed for all $x \in X$ then the set of points of strong continuity of $F$ is $\Pi^0_2$.

\item[(b)] If $Y$ is exhaustible by compact sets
and the set $F(x)$ is closed for all $x \in X$, then the set of
points of strong continuity of $F$ is $\Sigma^0_3$.
\end{enumerate}

\end{theorem}

We continue with the corresponding of Proposition \ref{proposition the set of points of continuity is analytic}.

\begin{prop}

\label{proposition strong continuity the set of points of continuity is co-analytic}

Let $(X,p)$ and $(Y,d)$ be complete and separable metric spaces
and let $F: X \Rightarrow Y$ be a multi-valued function such that the set $F \subseteq X \times Y$ is analytic. Then the set of points of strong continuity of $F$ is co-analytic.
\end{prop}

We conclude this article with some remarks which concern all previous results. The author would like to thank the anonymous referee for raising the questions stated below.

\begin{rem}

\normalfont

\label{remark about lightface}

All results above are in the context of classical descriptive set theory. One could ask whether the corresponding results are also true in the context of \emph{effective} descriptive set theory. In the latter context one deals with the notion of a recursive function $f: \om^k \to \om$ and of a recursive subset of $\om^k$. We assume that our given metric space $(X,d)$ is complete, separable and that there is a countable dense sequence \set{r_i}{i \in \om} such that the relations $d(r_i,r_j) < q$, $d(r_i,r_j) \leq q$ for $i,j \in \om$ and $q \in \mathbb{Q}^+$, are recursive. (An example of such space is $\mathbb{R}$ with $\set{r_i}{i \in \om}=\mathbb{Q}$.) One takes then the family \set{N(X,s)}{\s} of all open balls with centers from the set \set{r_i}{i \in \om} and rational radii and defines the class of \emph{semirecursive} sets or ``effectively open" sets as the sets which are recursive unions of sets of the form $N(X,s)$. The analogous notions go through the whole hierarchy of Borel and analytical sets i.e., one constructs the family of effectively closed, effectively $G_\delta$, effectively analytic sets and so on. The latter classes of sets are also called \emph{lightface} classes. The usual inclusion properties hold also for the lightface classes. For example every effectively closed set is effectively $G_\delta$. We should point out that there are only countably many subsets of a fixed space $X$ which belong to a specific lightface class. Also all singletons $\{q\}$ with $q \in \mathbb{Q}$ belong to every one of the lightface classes mentioned above except from the one of semi-recursive sets. The reader can refer to \cite{yiannis} for a detailed exposition of this theory. One natural question which arises is if the results which are presented in this article hold in the context of effective descriptive set theory. For example: if $F: \mathbb{R} \Rightarrow \mathbb{R}$ is a bounded multi-valued function such that the set $F(x)$ is effectively closed, is it true that the set of points of continuity of $F$ is effectively $G_\delta$? As the next proposition shows the answer to this question is negative even if $F$ is a single-valued function.
\end{rem}

Let us say that a family of sets \G \ is \emph{closed under negation} if whenever $A \subseteq X$ is in \G \ then $X \setminus A$ is in \G \ as well.

\begin{prop}

\label{proposition counterexample for effective}

Suppose that \G \ is a class of sets which is closed under negation and the family \newline \set{A\subseteq \mathbb{R}}{A \in \G} is countable. Then there is a function $f: \mathbb{R} \to \set{\frac{1}{n+1}}{\n} \cup \{0\}$ such that the set of points of continuity of $f$ is not a member of \G. In particular (by choosing \G \ as the lightface $\Delta^1_2$ class) there is a function $f: \mathbb{R} \to [0,1]$ such that the singleton $\{f(x)\}$ is effectively closed for all $x \in \mathbb{R}$ but the set of points of continuity of $f$ is neither effectively analytic nor effectively co-analytic.
\end{prop}

\textbf{Question 2.} In case we take $\G$ to be the lightface $\Delta^0_n$ class for some small \n, it would be interesting to see whether one can construct a function $f$ which satisfies the first conclusion of the previous proposition and has the additional property that the graph of $f$ belongs to \G.

\bibliographystyle{eptcs}

\end{document}